\documentstyle[amstex,preprint,aps]{revtex}

\begin{document}
\title{Transitions to Measure Synchronization in \\
Coupled Hamiltonian Systems}
\author{Xingang Wang$^{1}$ and Gang Hu$^{2,1}$}
\address{$^{1}$Department of physics, Beijing Normal University, Beijing\\
100875, China\\
$^{2}$China Center for Advanced Science and Technology (CCAST)\\
(World Laboratory), P.O. Box 8730, Beijing 100080, China}
\date{\today }
\maketitle

\begin{abstract}
Transitions to measure synchronization in two coupled $\phi ^{4}$ lattices
are investigated based on numerical simulations. The relationship between
measure synchronization (MS), phase locking and system's total energy is
studied both for periodic and chaotic states. Two different scalings are
discovered during the process to MS according to phase locking.\ Random walk
like phase synchronization in chaotic measure synchronization is found, and
phase locking interrupted by phase slips irregularly is also investigated.
Meanwhile, related analysis is qualitative given to explain this phenomenon.%
\end{abstract}

\newpage

The phenomenon of synchronization has been studied since 17th \cite{Huygens}%
, but the related research never stopped. Recently, in order to explore the
collective behavior in complex systems, this topic has been paied more
attention than ever before and became a hot focus in many science fields %
\cite{fields}. Although synchronization is first mentioned in periodic
systems, synchronization in chaotic systems attracted more attentions due to
its rich dynamical properties and potential applications in practical
systems \cite{pecora}. During the research process, the original concept has
been largely extended, many new phenomena had been discovered, phase
synchronization \cite{Rosenblum}, lag synchronization \cite{ls}, generalize
synchronization \cite{gs} had been the familiar words in new related works.
Till to now, most attentions had been concentrated on dissipative systems,
the related research for conservative system, the hamiltonian system, has
been neglected and only few works appeared \cite{H-syn}. Hamiltonian system
stands as the central topic of both classical and quantum mechanics, and a
large variety of practical systems can be well approximated by Hamiltonian
formalism even at weak dissipation. Thus, it is important to understand the
behavior of Hamiltonian systems in their possible synchronization process.

The former works abut Hamiltonian systems are mainly about their collective
behavior in globally coupled high dimension systems or studied their
transitions to equipartition from non-equilibrium state based on statistical
methods \cite{equipartation}, few paper had concerned to investigate their
detail dynamical behaviors during the system's evolutions. The reason mainly
because that Liouvill's theory request the phase space be conservative, it
prevents the full collapse of the orbits which related with complete
synchronization and phase synchronization, this means the classical concept
for synchronization developed in dissipative system is unsuitable for this
kind of system. But this situation has been changed in 1999, a new
phenomenon, measure synchronization (MS), for Hamiltonian systems is
discovered by Hampton and Zenette \cite{ms}. In their work, MS is defined as
all orbits cover the same region in the phase space with identical invariant
measures, and the transition to this state is characterized by a critical
coupling strength. Although MS has been found and critical coupling is
defined, that are still many questions still leaving unsolved. For example,
the route from desynchronization to MS is unclear, and how about system's
variables behave and what's the innate character for MS still need to be
cleared. Also, the relation between chaos and MS need to be considered
again, especially, the dynamical behavior for MS in chaotic state is also
valuable to study carefully. In this paper, we will concerned mainly on
these questions, and try to explore the way from non-MS to MS in coupled
Hamiltonian systems.

Let's consider two coupled classical $\phi ^{4}$ lattices, the system's
Hamiltonian reads \cite{phi-4}%
\begin{equation}
H=\frac{p_{1}^{2}+p_{2}^{2}}{2}+\frac{q_{1}^{4}+q_{2}^{4}}{4}+\frac{%
\varepsilon }{2}(q_{1}-q_{2})^{2}
\end{equation}%
the direct physical image is two coupled point masses each of them
oscillated in an quartic potential. Due to the quartic potential, system
could displays complex behavior and may express new property out of
intuition. By changing coupling strength $\varepsilon $ and initial values,
both period and chaos can be realized. The corresponding canonical equations
are%
\begin{eqnarray}
\stackrel{.}{q}_{1} &=&p_{1},\text{ \ \ \ \ \ }\stackrel{.}{p}%
_{1}=-q_{1}^{3}+\varepsilon (q_{2}-q_{1})  \nonumber \\
\stackrel{.}{q}_{2} &=&p_{2},\text{ \ \ \ \ \ }\stackrel{.}{p}%
_{2}=-q_{2}^{3}+\varepsilon (q_{1}-q_{2})
\end{eqnarray}%
In this model, the system's total energy is conserved, $E(t)=\frac{%
p_{1}^{2}(t)+p_{2}^{2}(t)}{2}+\frac{q_{1}^{4}(t)+q_{2}^{4}(t)}{4}+\frac{%
\varepsilon }{2}(q_{1}(t)-q_{2}(t))^{2}=E(0)=$Const, and initial values act
as an independent additional parameters. If we set each lattice's initial
positions at zero, then total energy will be independend to coupling
strength, this will be used in the following.

Once two lattices reached MS, then any macroscopic values will equal if only
the average time long enough. One example is the energy of each lattice,
defined as%
\begin{equation}
e_{i}=\frac{1}{T}\int\nolimits_{0}^{T}[p_{i}^{2}(t)/2+q_{i}^{4}(t)/4]dt,%
\text{ \ \ \ }i=1,2
\end{equation}%
before MS, two orbits occupy different domains which never intersect, of
course $e_{1}\neq e_{2}$, this situation continued till MS reached, after
the critical coupling, two lattice's energy be equaled due to their
identical invariant measures. In Fig. 1 we choose initial condition%
\begin{eqnarray}
q_{1} &=&0,\text{ \ }q_{2}=0,  \nonumber \\
p_{1} &=&0.1,\text{ }p_{2}=0.2
\end{eqnarray}%
and plotted each lattice's energy $e_{i}$ versus $\varepsilon $. During our
numerical simulations, third order bilateral symplectic algorithm \cite{bsa}
and small time step $h=0.001\,$\ are used to keep the system's conservative
property. The critical coupling, $\varepsilon _{c}=0.0032,$ which separate
non-MS and MS is clearly shown in this figure. Meanwhile, we also plotted
two lattice's total energy $e=e_{1}+e_{2}$ versus $\varepsilon $ in the same
figure for comparison. It is shown that energy $e$ has its minimum on the
critical point $\varepsilon _{c}$, below and above this point $e$ decreased
and increased monotonically. This made us reminisced phase synchronization
which has been warmly studied recently \cite{Rosenblum}\cite{phaseslip-1}%
\cite{phaseslip-2}, made it interesting to investigate the phase locking, or
named phase synchronization, in this system. In Hampton's work, the phase
difference $\Delta \theta =\theta _{1}-\theta _{2},$ with $\theta _{1,2}$
denotes each orbit's angle and defined on the real line rather than on the
circle, between two orbits is found to obey different rules. Before MS, $%
\Delta \theta $ increases linearly with time goes on, and above MS, the
difference oscillates with a characteristic frequency which around a
constant value. But only these can't explain the minimum, the deep reason
need to be explored.

More information about MS and phase locking can be found in Fig. 2, in Fig.
2(a) and (b) we adopted $\varepsilon =0.002<\varepsilon _{c},$ two orbits
stay in different domain, and phase difference increases linearly with speed 
$\eta =0.1$. While $\varepsilon =0.0035>\varepsilon _{c},$ MS reached and
phase difference is oscillates within $(-\pi ,\pi )$ with frequency $f=2.12$%
, as what is shown in Fig. 2(c) and (d). In order to explore the relation
between phase difference and coupling strength, we plot $\eta $, which is
defined as the speed of phase difference between two orbits before MS, and $%
f,$ which is defined as the oscillate frequency of phase difference after
MS, versus $\varepsilon $ in Fig. 2(e) and (f). It is observed that, before
MS, $\eta $ decreases gradually as $\varepsilon $ increases and trend to
zero on $\varepsilon _{c}$, while $f$ increases gradually from zero as $%
\varepsilon $ increases from $\varepsilon _{c}$, this is coincide to Ref. %
\cite{ms}.

During our numerical simulations, we found that phase locking exists two
different trends corresponding to non-MS and MS respectively, future study
shows that its right the reason why the energy $e$ show its minimum at $%
\varepsilon _{c}.$ Before the critical point, MS doesn't exist and orbits
have different main frequencies, but we found that: except $\eta $ decreases
as $\varepsilon $ increases, the distribution of phase difference $\varphi $
(model $\Delta \theta $ with $2\pi $) has a obviously trend. In Fig. 3(a) we
plotted the distributions for $\varphi $ under three different coupling
which all below the critical point. From this figure, we can find that as $%
\varepsilon $ increases, the probability for $\varphi =\pi $ changes to
higher, we remark here that $\pi $ is the biggest phase difference for two
orbits after modeled $\theta _{1}$ and $\theta _{2}$ with $2\pi .$ In Fig.
3(b) we also plotted the probability distributions for distance difference $%
\Delta q=|q_{1}-q_{2}|,$ it is found that with $\varphi $ trends to $\pi $
in distribution, the probability for large $\Delta q$ increases as well.
After $\epsilon >\varepsilon _{c},$ as coupling increases, the oscillate
frequency rise and the amplitude decrease, meanwhile, the distribution for $%
\Delta q$ contracted to small scope.

Already had these ideas in mind, the reason for the minimum of energy $e$ on
the critical point is straightforward. As the coupling increases, the
relation between two phases become stronger, although they are unlocked, but
they have the trend to reach phase locking, this can be found from $\eta $
decreases as $\varepsilon $ increase. As we already known, for Hamiltonian
system, the classical phase synchronization can't be reached due to its
conservative character, but as coupling increases, two orbits trend to reach
it. On the way to phase synchronization, the first $\Delta \theta $ should
be $\pi $, because $\pi $ stands for the largest phase difference if we
model phase by $2\pi .$ This directly made the probability of large $\Delta
q $ increased and thus induced energy $e$ decrease in this process. Advanced
increase the coupling, $\varepsilon >\varepsilon _{c},$ then two orbit
become MS and be phase locked, under this situation, the coupling strength
is strong enough to reach MS, but is weaker to bind the phase oscillate in a
small amplitude, continue increase $\varepsilon $, the amplitude will drops
from large value $\pi $, where $\varepsilon $ slightly larger than $%
\varepsilon _{c},$ to zero. During this process, phases are locked and the
oscillate amplitude decreased, induced the distribution for $\Delta q$
inclined to short distance, at last made energy $e$ tend upward. In order to
give a more convincing expression, we plot the probability of $\pi $ (denote
by $\rho _{\pi }$) versus $\varepsilon $ in Fig. 4(a), $\log (\rho _{\pi })$
for vertical axis and $\log (\varepsilon _{c}-\varepsilon )$ for horizontal
axis. A fine scaling is found%
\begin{equation}
\rho _{\pi }=\frac{\chi }{(\varepsilon _{c}-\varepsilon )^{\nu }}
\end{equation}%
with $\nu =1.96\approx 2.0$ and $\chi $ denotes as constant. Also, in Fig.
4(b) we plotted amplitude $A$ versus $\varepsilon $ in MS region, this time
the scaling is found to be a exponential mode%
\begin{equation}
A=A_{0}+Be^{-\frac{\varepsilon }{\alpha }}
\end{equation}%
with $\alpha =1.0\times 10^{-3}$ and $A_{0},$ $B$ constants.

Under the situation mentioned above, $\varepsilon _{c}$ not only stands as
the critical point for MS and the energy minimum, but also stands as the
critical point for two different trends of phase locking, i.e., before MS,
two orbits are trend to achieve the first possible locking $\Delta \theta
=\pi ,$ after MS, the amplitude of oscillation dropped from $\pi $ and trend
to approach zero. But we should also noticed here that all what we have
discussed is just within period, no chaos appeared, so a natural question is
how about these phenomena within chaos, does different phase locking trend
still exist, and what's the new characters for MS when system is chaos. All
these questions aroused our interests to investigate MS in coupled chaotic
Hamiltonoian systems.

In order to get chaos, this time we choose initial condition%
\begin{eqnarray}
q_{1} &=&0,\text{ }q_{2}=0,  \nonumber \\
p_{1} &=&0.1,\text{ }p_{2}=0.4
\end{eqnarray}%
Similar to Fig. 1, we plotted energy $e_{i}$ and total energy $e$ versus
coupling $\varepsilon $ in Fig. 5(a), different to the above result, energy $%
e_{i}$ behaves a disordered way in chaotic regions, made it difficult to get
the critical point for MS. Through tracing the time evolution of their
orbits, we found that the uneven behavior of $e_{i}$ is closely related to
chaos, in other word, its just the conjoint action between chaos and MS that
lead to this irregular behavior. On one hand, the definition of MS requests
each lattice's energy be equaled, just as what happened in periodic state,
but on the other hand, the internal stochastic character keep their orbits
moving in an irregular way. Thus, $e_{1}=e_{2}$ can be reached only under
limit $T\rightarrow \infty .$ However, we found the behavior of summation $e$
is more regular than $e_{i}$, despite of some abrupt windows. Through
calculate system's largest Lyapunov exponent $\lambda $, it is found that
these windows are corresponding to period which embed in chaos. In Fig. 5(b)
we plotted $e$ and $\lambda $ versus $\varepsilon $ in the same figure,
compare with Fig. 5(a), it is shown that $e$ decreases more smoothly then $%
e_{i}$ in chaotic regions. This result is somewhat similar to the behavior
of $e$ in region $\varepsilon <\varepsilon _{c}$ plotted in Fig. 1. As we
have known, the trends of phase locking plays important role for the minimum
of $e$ in periodic state, so we may ask does these the same to chaos, if
not, what's the reason made $e$ decrease in chaotic state and how about the
phase locking now?

On important difference between this work and Hampton's is, in our model,
chaos is closely related with MS, and the largest Lyapunov exponent is a
suitable order parameter for MS. We choose $\varepsilon
=0.025>0.012=\varepsilon _{c}$, with initial condition Eqs. 7, system is
chaos with $\lambda =0.08$. Phase difference $\Delta \theta $ versus time $t$
under this coupling strength is plotted in Fig. 6. It is shown that $\Delta
\theta $ behaves a random walk process around zero, and there also exist
some platforms with different wide. The inset figure is enlargement of the
rectangle region in original figure, there the details about the platform is
shown. In fact, each platform is in the state of phase locking, in these
regions, although $\Delta \theta $ oscillate all the time, but its amplitude
is small than $\pi $. We also found that even in regions where phase are
unlocked, the phase difference shows a $2\pi $ like slips. The same
phenomenon also can be found in others chaotic states with different $%
\varepsilon >\varepsilon _{c}$, the only difference is, as $\varepsilon $
increases, the random walk more near zero, and more time two orbits be phase
locked.

The $2\pi $ slips in phase locking made us reminisced the phase slips in
phase synchronization of dissipative systems \cite{phaseslip-1}\cite%
{phaseslip-2} where various models were used and rich phenomena were
discovered. But in our opinion, these results are unsuitable to Hamiltonian
system due to its conservative property, here we only give a qualitative
analysis for this phenomenon without proofs. When system is MS and within
chaotic state, the coupling is stronger and trend to lock phases, this is
exhibited by the platforms in Fig. 6(a), but due to the stochastic property
of chaos, made the locking unstable, this is exhibited by the phase slips.
Because the first possible locking is near $\pi ,$ and sometimes $\pi $
locking is unstable, made $\varphi $ stays a long time within $(0,\pi )$
(phase locking), then broke suddenly because the stochastic property of
chaos, after $2\pi $ increase, $\varphi $ travels again near $\pi ,$ this
time it has two choices, continue this slip if its still local unstable, or
stay on locking state except local stable be changed. In our studies, we
found that, in chaotic state, although phase locking doesn't exist, but $%
\varphi $ still owns the trend to be $\pi $ locked as $\varepsilon $
increases, and the distribution for $\Delta q$ also incline toward long
distance, these properties made $e$ decreased in chaotic state with $%
\varepsilon $ increases.

In conclusion, we have investigated the way from non-MS to MS in coupled $%
\phi ^{4}$ lattices both for period and chaos. Due to Hamiltinian system's
conservative property, exactly phase locking, $\Delta \theta =0\,$, is
unreachable, but we found that, in periodic state, $\varphi $ trend to $\pi $
as $\varepsilon $ approach $\varepsilon _{c}$, after $\varepsilon
>\varepsilon _{c},$ $\Delta \theta $ is oscillate around zero and the
related amplitude decreases with $\varepsilon $ increases, on the critical
point $\varepsilon _{c}$, energy $e$ own its minimum, also, different
scalings has been found to describe these trends. MS is found to be
connected closely with chaos in this model, random walk like phase locking
is discovered, after analysed system's dynamical behaviors, we gave the
qualitative explaining for this phenomenon. We argued that in some models,
like $\phi ^{4},$ Lyapunov exponent can be a suitable parameter for MS. But
there are still some works ahead waiting for future research. The first work
is to explain the phase locking and slip in chaotic MS strictly, we noticed
that a recent work about phase synchronization also shown a random walk like
phase locking \cite{phaseslip-2}, its gave an example of nontrivial
interaction between microscales and macroscales, so another work is to
investigate the scalings for phase locking in chaotic states, also, to study
these phenomena in large systems in also a interesting work. We wish this
paper could deepen people's understanding of MS, and the observed phenomenon
could potentially be investigated in many other related research in coupled
Hamiltonian systems, e.g., elliptical galaxies, Hamiltonian version
Kuramoto's model, transitions to equipartition in coupled large Hamiltonian
systems, and we also hope this work could helpful in the research of quantum
chaos.

This research was supported by the National Natural Science Foundation of
China, the Nonlinear Science Project of China, and the Foundation of
Doctoral training of Educational Bureau of China.

Captions of figures

Fig. 1 The time average energy for each lattice $e_{i}=%
\mathrel{\mathop{\lim }\limits_{T\rightarrow \infty }}%
\frac{1}{T}\int_{0}^{T}(\frac{p_{i}^{2}}{2}+\frac{q_{i}^{2}}{4})dt,$ $i=1,2$
versus coupling strength $\varepsilon ,$ critical coupling for MS is $%
\varepsilon _{c}=0.0032.$ Total energy $e=e_{1}+e_{2}$ is plotted on the
left vertical axis (denoted by hollow circle), on the critical point, $e$
has its minimum, before and above $\varepsilon _{c}$, $e$ decreases and
increases monotonically.

Fig. 2 With the initial condition Eqs. 4, when $\varepsilon
=0.002<\varepsilon _{c}$, (a) plot the phase difference $\Delta \theta $
versus $t$, the slope rate $\eta =0.1,$ (b) plot phase orbits for two
lattices, it is shown that two orbits moving in different domains without
overlap. (c)\ and (d)\ corresponding to $\varepsilon =0.0035>\varepsilon
_{c} $, $\Delta \theta $ oscillate with frequency $f=2.12,$ and two orbits
be MS. (e) plote $\eta $ versus $\varepsilon $ for $\varepsilon <\varepsilon
_{c}$ and (f) plot $f$ versus $\varepsilon $ for $\varepsilon >\varepsilon
_{c}$.

Fig. 3 (a) Probability distributions of $\varphi $ for three different $%
\varepsilon $s, circle for $\varepsilon =0.001,$ upward triangle for $%
\varepsilon =0.002$ and downward triangle for $\varepsilon =0.025$, all
below $\varepsilon _{c}.$ (b)\ Probability distributions for distance
difference $\Delta q,$ the couplings same to (a).

Fig. 4 (a) The $\pi $ distribution of $\varphi $ versus coupling for $%
\varepsilon <\varepsilon _{c}$, after logarithm, all data lie on one line,
the related scaling reads $\rho _{\pi }=\frac{\chi }{(\varepsilon
_{c}-\varepsilon )^{\nu }}$ with $\gamma =1.96.$ (b)\ Oscillation amplitude $%
A$ versus $\varepsilon ,$ scaling relation reads $A=A_{0}+Be^{-\frac{%
\varepsilon }{\alpha }}$ with $\alpha =1\times 10^{-3}.$

Fig. 5 Initial condition as Eqs. 7. (a) The summational energy $e$ and each
lattice's energy $e_{i}$ versus $\varepsilon $, hollow circle represents the
summation. (b)\ $e$ and the largest Lyapunov exponent versus $\varepsilon .$
It can be found that in chaotic MS regions the summation decreases as $%
\varepsilon $ increases.

Fig. 6 $\varepsilon =0.025,$ the time evolution of phase locking in chaotic
MS, $\lambda =0.08.$ The inset window is enlargement of the rectangle region
in original figure, phase locking and slips can be found.

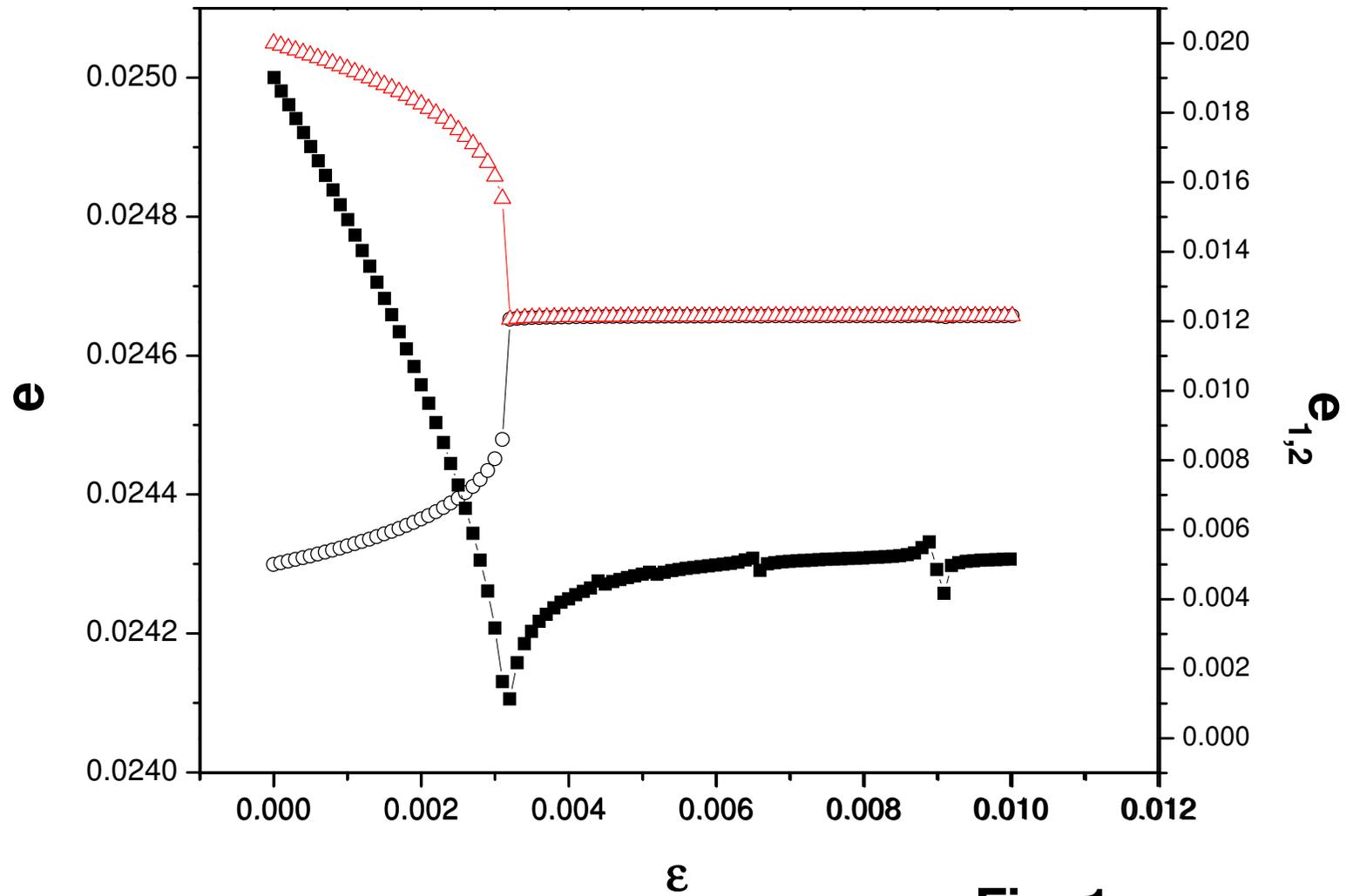
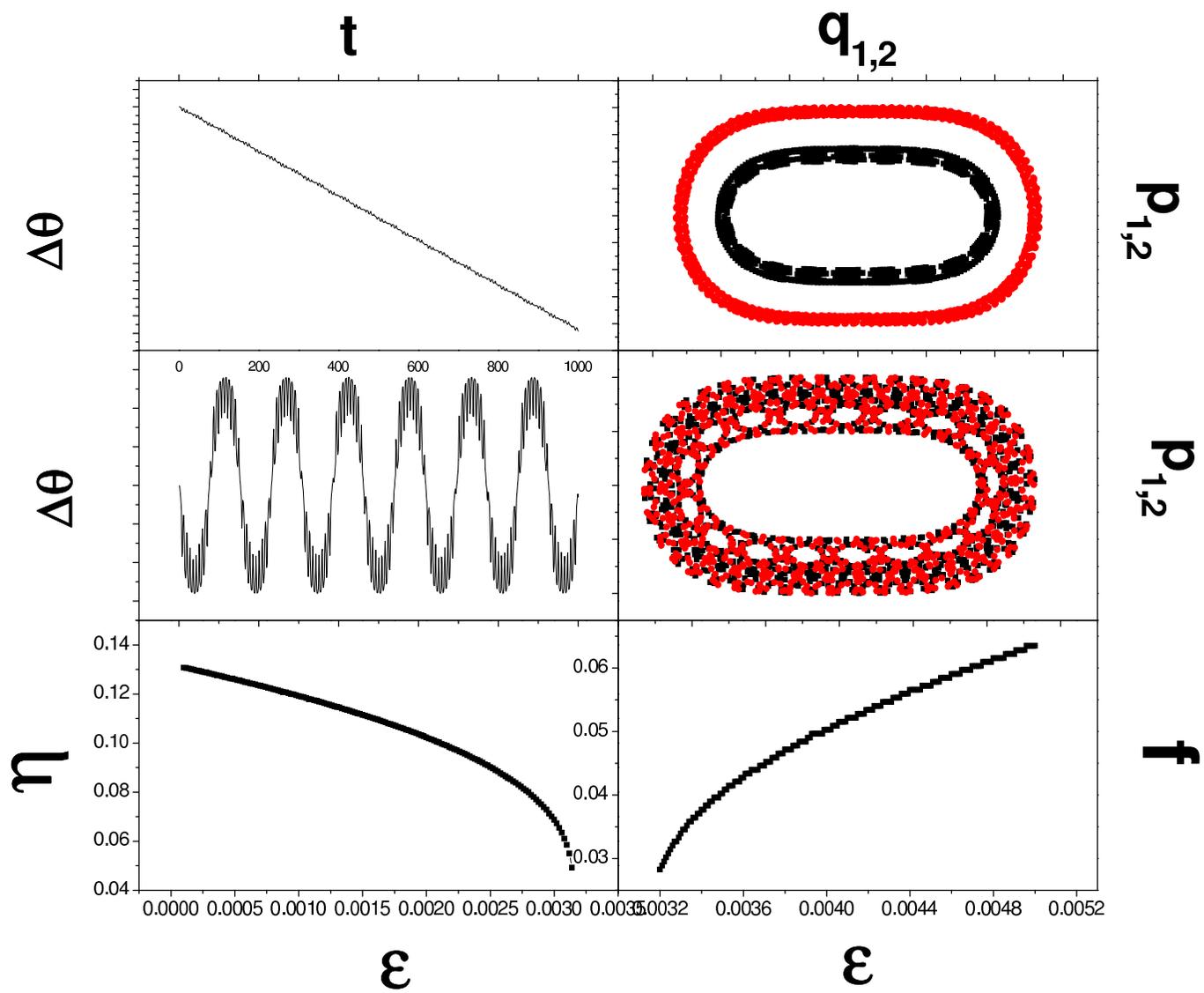
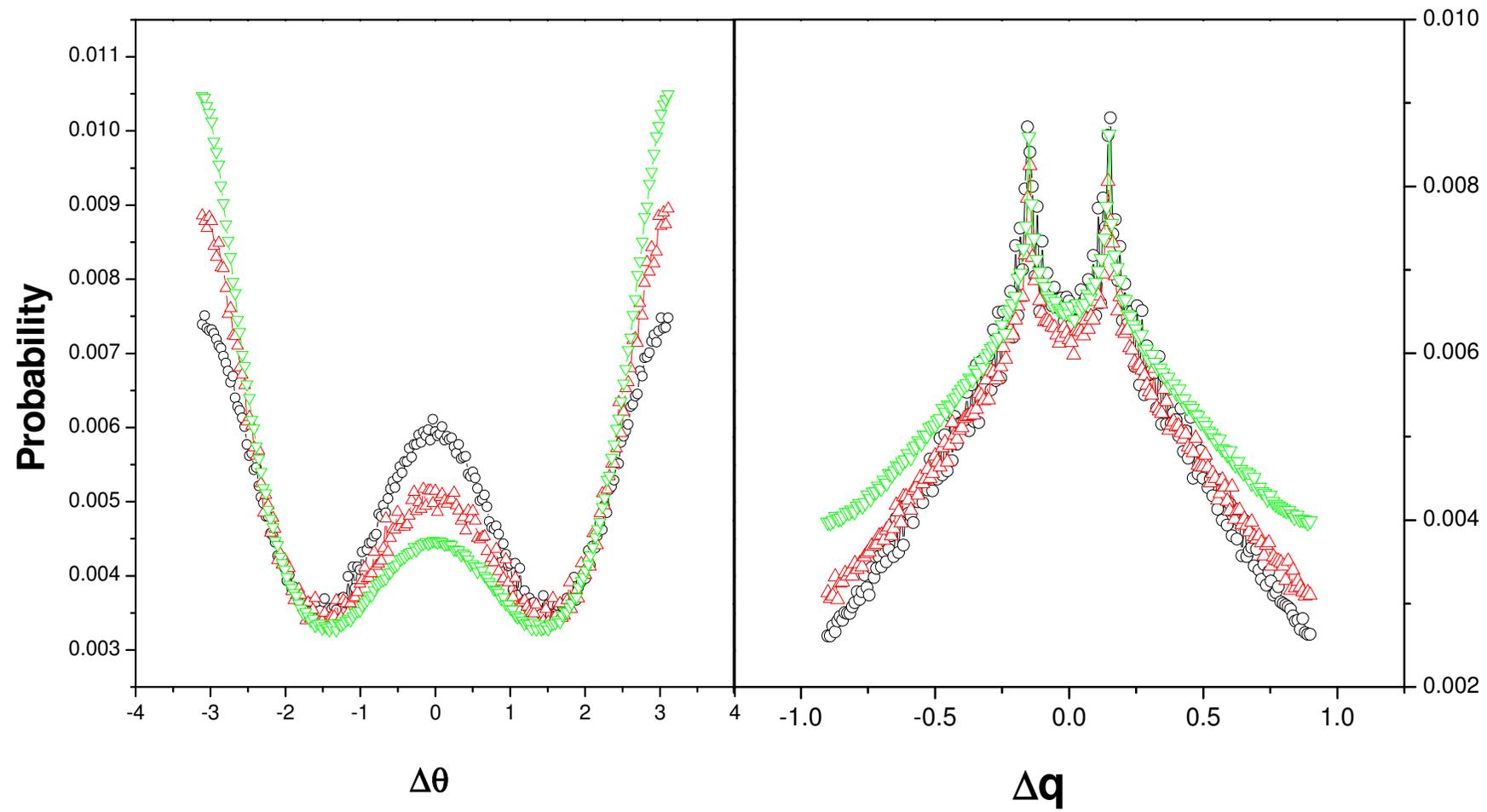
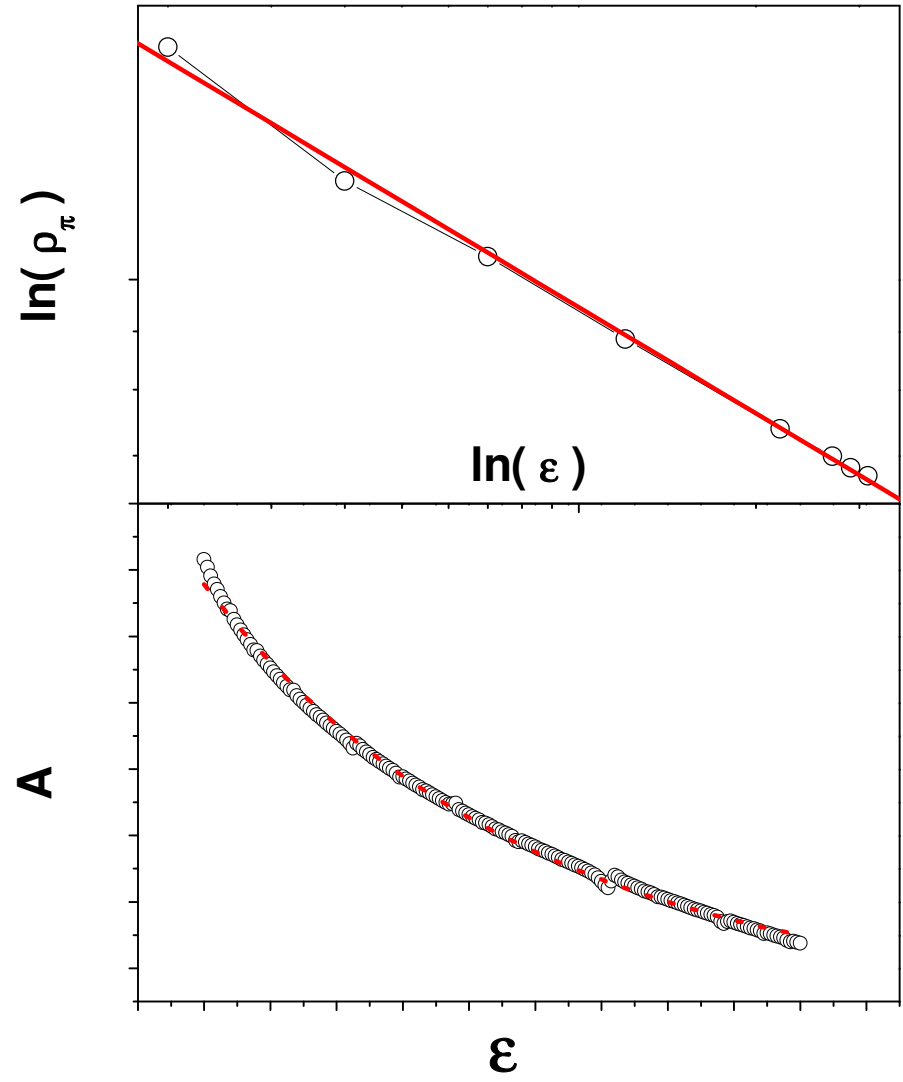
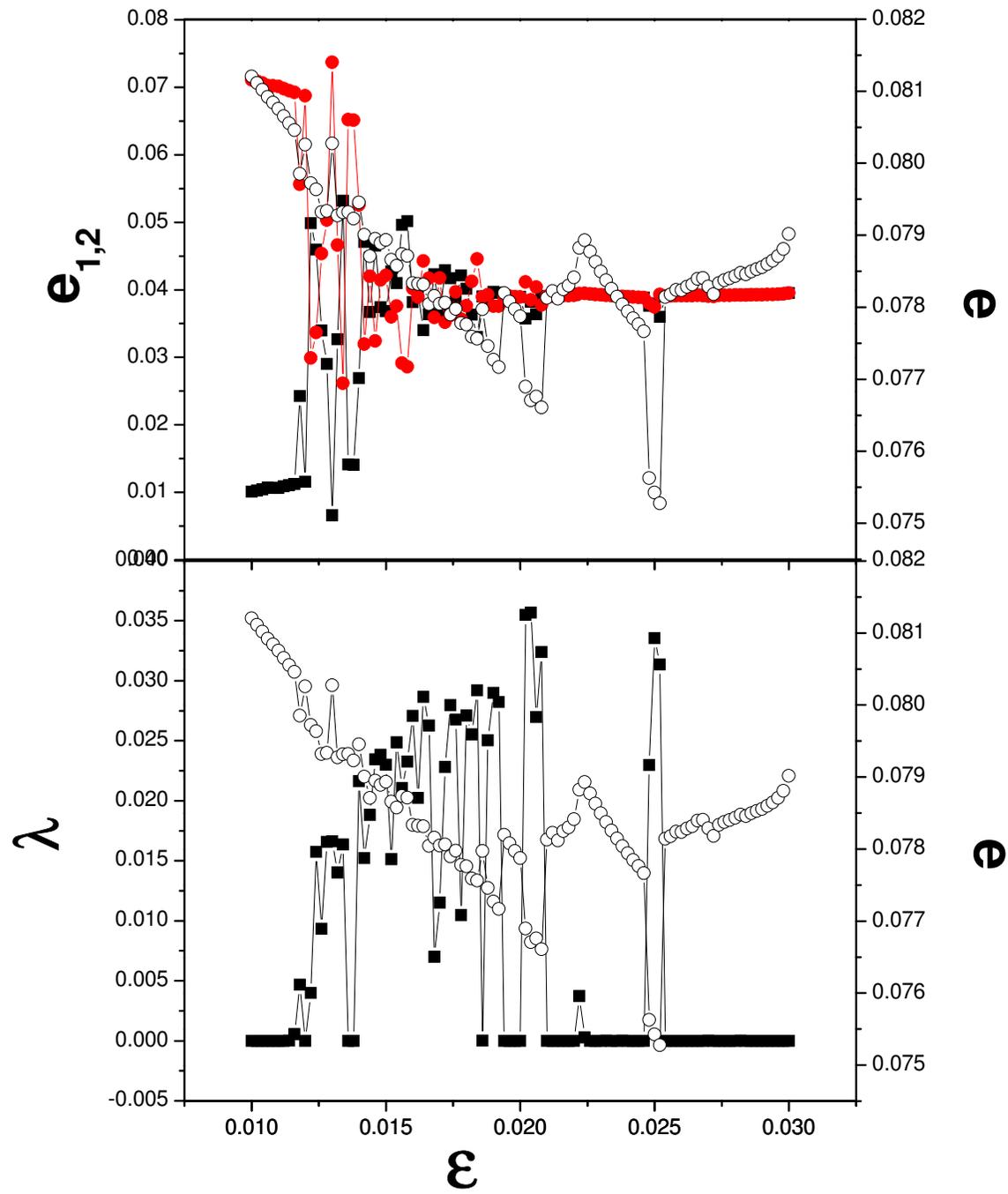
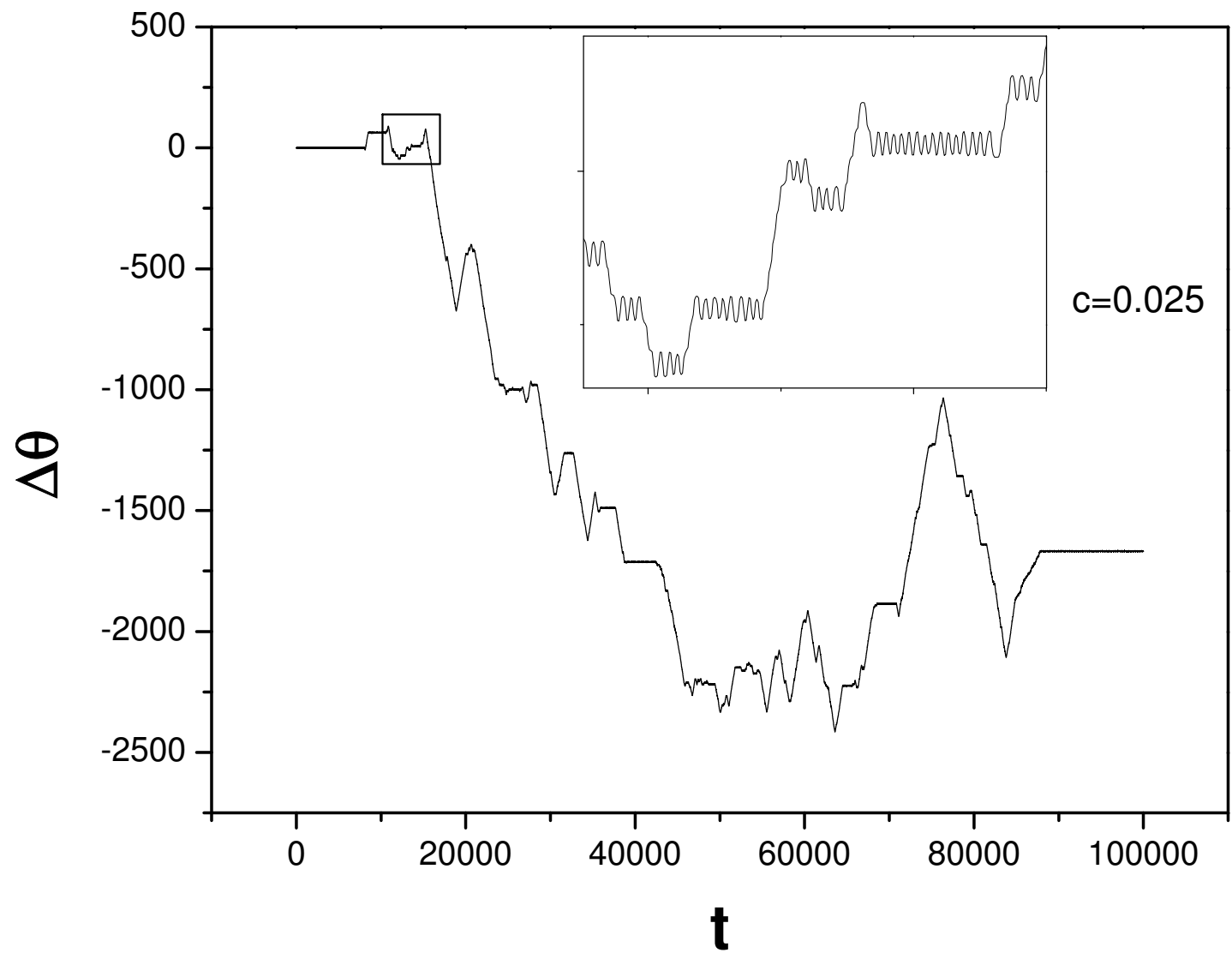
\end{document}